\documentclass[12pt, a4paper]{iopart}
\usepackage{iopams}
\usepackage{graphicx}
\begin{document}
\title{Heavy quark thermodynamics in full QCD}
\author{Konstantin Petrov (for the RBC-Bielefeld Collaboration)}
\address {Niels Bohr Institute, Blegdamsvej 17, 2100 Copenhagen O, Denmark\\ and SDU, Odense, Denmark}
\ead{kpetrov@nbi.dk}
\begin{abstract}We analyze the large-distance behaviour of static 
quark-anti-quark pair correlations in 
QCD.  The singlet free energy is calculated and the entropy contribution to it 
is identified allowing us to calculate the excess internal energy. The free 
energy has a sharp drop in the critical region, leading to sharp peaks in both 
excess entropy and internal energy. 
\end{abstract}
\pacs{  11.15.Ha,  11.10.Wx, 12.38.Mh, 25.75.Nq} 
\section{Static quark anti-quark free energies}
Among its many other applications, finite temperature lattice QCD is a useful 
tool to study  the modification of quark-anti-quark forces n a hot medium. 
Recent progress in large-scale simulations allowed us to address this problem 
in QCD with realistic quark masses. This study is based on a ongoing 
RBC-Bielefeld 
joint research effort which is fully described in Ref.\cite{us}. 
Here we will note that calculations are
performed  at physical strange quark mass and two light quark masses 
$m_q=0.1m_s$ and $0.2m_s$, with $m_s$ being the strange quark mass; 
two cutoffs, corresponding to temporal extent $N_t=4,6$ were used.

We discuss here the asymptotic large distance behavior of 
$\bar{q}q$ renormalized free energies. This is a gauge invariant observable.
However, it is most suitable to extract it from a $\bar{q}q$ singlet free 
energy, which
is gauge variant and thus has been calculated by us in Coulomb 
gauge \cite{ophil02,okacz02,digal03,okaczlat03,okacz04,petrov04,okacz05}.
To extract singlet contribution to the free energy one uses a correlation 
function of 
two static quarks, represented by temporal Wilson lines, one 
representing a static quark at $x=0$ and another - a static anti-quark at 
distance~$r$,

\begin{equation}
\exp(-F_1(r,T)/T+C)= \frac{1}{3} \langle Tr W(\vec{r}) W^{\dagger}(0) \rangle 
\; .
\label{f1def}
\end{equation}
This is, of course, not the full free energy of the system but the excess 
free energy which is generated by adding a $\bar{q}{q}$-pair to the hot bath.
In the zero temperature limit the singlet free energy defined above coincides 
with the well known static potential, as demonstrated in e.g. 
Ref. \cite{milc04}.  At finite 
temperature $F_1(r,T)$ gives information about in-medium modification of 
inter-quark forces and color
screening. Both the singlet free energy and  the zero temperature static 
potential are 
defined up to an additive constant $C$ which depends on the lattice cutoff.  
Our renormalization scheme consists of using the zero temperature potential 
as normalization at short distances, where the effect of the medium is 
negligible. In the transition region we used a Cornell-type ansatz for the 
potential,

\begin{equation}
V(r)=-\frac{0.385}{r}+\frac{1.263}{r_0^2} r \;.
\label{t0pot}
\end{equation}
which gives a good parametrization of the lattice data for the zero temperature
static potential calculated in Ref.\cite{us}.  At much higher temperatures, 
however, as we probe the free energy at increasingly  shorter distances, where 
the effects of the running coupling can become important, we calculated $V(r)$ 
explicitly at every temperature and used them for normalization.
Our results for the singlet free energy and the scaling study have been 
presented 
in Ref.\cite{lat06}. Here we will focus on the large-distance quantities.

\section{Renormalized Polyakov Loop}

The expectation value of the Polyakov loop, the trace of temporal Wilson 
line, $\langle L(\vec{r}) \rangle =\langle Tr W(\vec{r}) \rangle$,
is the order parameter
for the deconfining transition in pure gauge theories. In 
QCD dynamical quarks the relevant $Z(3)$ symmetry is broken explicitly and 
 $\langle L(\vec{r}) \rangle$ thus  is no longer an order parameter. 
Still it remains an interesting quantity to study the deconfinement 
transition as it shows a rapid increase in the
crossover region \cite{us,milcthermo,fodor06} and is one of the ways which 
can be used to determine the transition temperature \cite{us,fodor06}.
The unrenormalized Polyakov loop defined above strongly depends on the lattice 
cutoff and has 
no meaningful continuum limit. On the other hand a correlator of Polyakov loops is a physical quantity and  corresponds to the color averaged free energy up to a normalization constant.
Due to the cluster decomposition of correlation functions we have,
\begin{equation}
\exp(-F(r,T)+C)=\frac{1}{9} \langle L(\vec{r}) L^{\dagger}(0) \rangle|_{r \rightarrow \infty}= |\langle L (0)\rangle|^2.
\end{equation}
Here the normalization constant $C$ is the same as in Eq.\ref{f1def}. Moreover, 
at large distances the relative color orientation of the $\bar{q}q$-pair
is not important and thus the color singlet free energy and the color averaged 
free energy approach the same constant value $F_{\infty}(T)$.
\begin{figure}
\includegraphics[width=8.3cm]{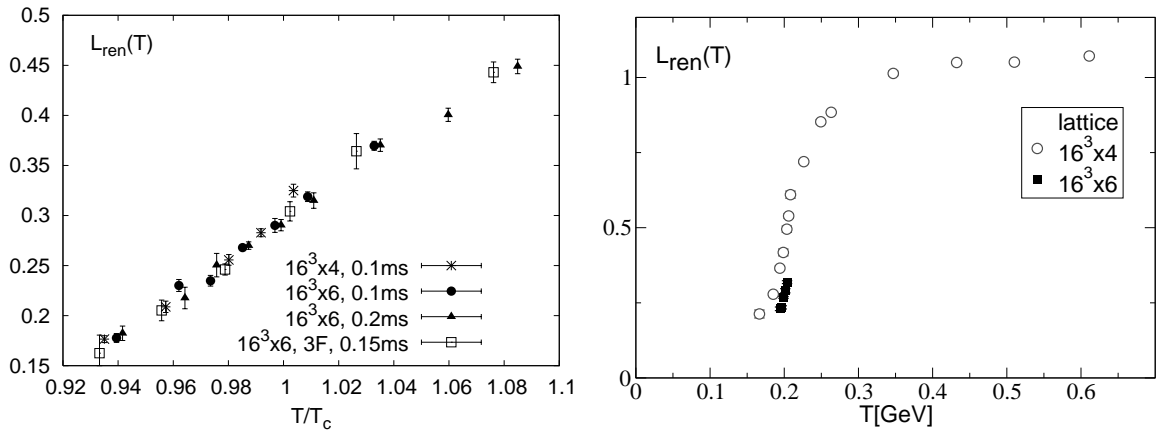}
\includegraphics[width=7.3cm]{LrenGev.eps}
\caption{
The renormalized Polyakov in the vicinity of the transition calculated on $16^3 \times 6$, $16^3 \times 4$ lattices
for different quark masses (left) and in the entire temperature range (right)}
\label{fig:lren_sum}
\end{figure}
Therefore, following Ref.~\cite{okacz02}, we define the renormalized 
Polyakov loop as 
\begin{equation}
L_{ren}(T)=\exp(-\frac{F_{\infty}(T)}{2 T}).
\end{equation}
First, in order to check the scaling behaviour, let us concentrate on the 
transition region. 
As one can see from Fig.\ref{fig:lren_sum} $L_{ren}(T)$ shows an almost 
universal behavior as function of $T/T_c$ for all quark masses studied by us, 
including the case of three degenerate flavors. This suggests that in the 
region of the small quark masses studied here, the flavor and quark mass 
dependence of the free energy can almost entirely be absorbed in 
the flavor and quark mass dependence of the transition temperature $T_c$. 
Note that the results obtained on $16^3 \times 4$ lattice are in remarkably 
good agreement with the results obtained on $16^3 \times 6$ lattices, 
indicating again that cutoff effects are small. For comparison with 
results obtained using different simulation techniques we refer to 
Ref.\cite{lat06}.
For higher temperatures we simulated a large ensemble of $16^3 \times 4$ 
(finite temperature). Here results obtained for the static potential at $T=$
on $16^3 \times 32$ (zero temperature) lattices where
used for the normalization. Results for the renormalized 
Polyakov loop are presented on Fig.\ref{fig:lren_sum} (right). As one can see 
it shows a rapid change in the temperature region from approximately 
180MeV to 230MeV and almost reaches a plateau at about 400MeV.   

\section{Entropy and Internal Energy}
\begin{figure}
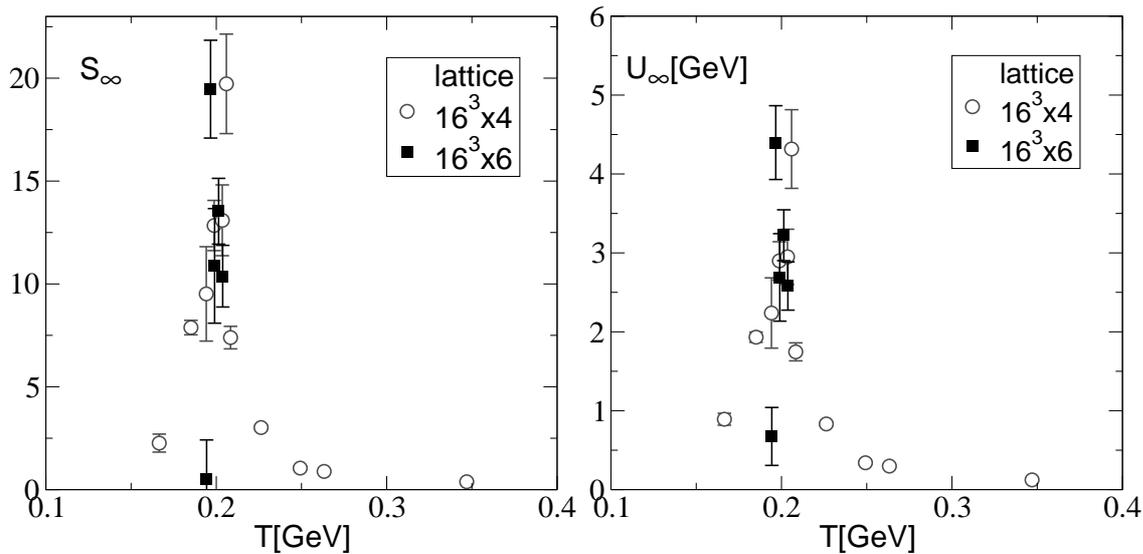

\includegraphics[width=7.5cm]{S.eps}
\includegraphics[width=7.3cm]{Uint.eps}
\label{fig:thermo}
\caption{Excess entropy and internal energy of the quark-anti-quark pair 
calculated on $16^3 \times 4$ lattices for $m_q=0.1m_s$}
\end{figure}
The excess free energy
obviously has a behaviour similar to that of the renormalized Polyakov loop 
through which it was defined. Its variation with temperature defines the
excess energy and entropy, respectively, derived from standard thermodynamic 
relations. As one can see from Fig.\ref{fig:thermo} the entropy 
and internal energy have sharp peaks in the transition region.
This can be intuitively understood in a following way: while QCD with
non-zero quark masses does not undergo 
phase transition the crossover is rapid enough to share some important 
properties, such as enhancement of the long distance correlations. Therefore, 
adding a $\bar{q}q$-pair to such a system would require very large, possibly 
overlapping, gluonic clouds to screen the color charges, which, in turn, 
gives rise to the aforementioned peak in the internal energy.

As a phenomenological consequences of this calculation we note that potential 
models which make use of our data(Refs.\cite{Mocsy:2004bv},\cite{Mocsy:2005qw})
fail to reproduce more rigorous results obtained in Ref.\cite{spectr}, namely 
charmonia survival till about $2.5T_c$ with simultaneous dissolution of 
the $\chi_c$. This shows the need for more refined phenomenological approaches.

\section{Conclusions}
We have calculated the singlet free energy of a static $\bar{q}q$-pair in QCD 
with realistic quark masses. We have found that this and related  quantities 
show little cut-off dependence and can be calculated reliably on relatively 
coarse lattices.  
The renormalized Polyakov loop derived from the large distance limit of the 
singlet
free energy has been calculated and shown to be in the scaling regime. The free 
energy experiences rapid drop in the transition region, which leads to
sharp peaks in the excess entropy and internal energy.

\end{document}